# Impact of Raman scattered noise from multiple telecom channels on fiber-optic quantum key distribution systems

Thiago Ferreira da Silva, Guilherme B. Xavier, Guilherme P. Temporão,
and Jean Pierre von der Weid, *Member, IEEE*

*Abstract*—In this paper we analyze the impact of the spontaneous Raman scattered noise generated from multiple optical classical channels on a single quantum key distribution channel, all within the telecom C-band. We experimentally measure the noise generated from up to 14 continuous lasers with different wavelengths using the dense wavelength division multiplexing (DWDM) standard, in both propagation directions in respect to the QKD channel, over different standard SMF-28 fiber lengths. We then simulate the expected secure key generation rate for a decoy-states-based system as a function of distance under the presence of simultaneous telecom traffic with different modulation techniques, and show a severe penalty growing with the number of classical channels present. Our results show that, for in-band coexistence, the telecom channels should be distributed as close as possible from the quantum channel to avoid the Raman noise peaks. Operation far from the zero dispersion wavelength of the fiber is also beneficial as it greatly reduces the generation of four-wave mixing inside the quantum channel. Furthermore, narrow spectral filtering on the quantum channels is required due to the harsh limitations of performing QKD under real telecom environments, with the quantum and several classical channels coexisting in the same ITU-T C-band.

## I. Introduction

Quantum key distribution (QKD) enables the generation of a secret key between two remote parties with security guaranteed by the principles of quantum physics [1]. It can be of great interest to the telecommunication industry as it can provide an important alternative to the key distribution problem in classical cryptography. The widespread practical deployment of QKD depends heavily on its compatibility with current telecom optical networks. From a practical point of view it would be highly desirable to have QKD channels sharing optical fibers together with telecom data channels. Although many QKD demonstrations have been done in "dark" fibers, that is, fibers devoid of any classical signals, there has been considerable interest in experimentally investigating QKD performed in fibers with co-existing classical signals [2-9]. This has huge benefits in order to minimize the cost of having an entire fiber solely dedicated to a QKD system.

Typically the classical and quantum signals are multiplexed using standard wavelength division multiplexing technology (WDM). Two main technical difficulties arise when trying to multiplex the two types of signals in the same optical fiber. The first one is related to the fact that the power level difference between a classical signal and the single-photon level can reach 100 dB, which may lead to extreme crosstalk and even eventually saturating the single-photon detectors. Cascaded filters, combined with pre-filtering of the classical channels to remove unwanted broadband spontaneous emission, is effective in dealing with this issue. The other hurdle is noise generated from the classical channels when photons are inelastically scattered due to the spontaneous Raman scattering (SRS) [4,10-15]. In this case, the problem is more complex as the noise is generated in-band with the QKD signal along the fiber, therefore it cannot be spectrally filtered out. Different solutions have been proposed such as: operating in the 1300 nm window outside of the Raman bandwidth [4], using reduced launch powers for the classical channels and narrowband filters [11], time interleaving the single photons with the Raman scattered photons [8] and employing a temporal filter [9].

In this work, the impact of spontaneous Raman noise generated from multiple classical DWDM channels on a QKD wavelength, all within the telecom C-band, is experimentally measured. In this scenario the secret key rate for a QKD system employing decoy-states is analyzed for the first time. This is performed in two configurations, co- and counter-

This work was supported by the Brazilian agencies CAPES, CNPq and FAPERJ. The work of G. B. Xavier acknowledges support of CONICYT PFB08-024, Milenio P10-030-F and FONDECYT no. 11110115.

T. Ferreira da Silva is with the Center for Telecommunication Studies, Pontifical Catholic University of Rio de Janeiro, Rio de Janeiro, RJ, Brazil. He is also with the Optical Metrology Division, National Institute of Metrology, Quality and Technology, Duque de Caxias, RJ, Brazil (e-mail: thiago@opto.cetuc.puc-rio.br).

G. B. Xavier is with the Departamento de Ingeniería Eléctrica, Universidad de Concepción, Concepción, Chile. He is also with the Centre for Optics and Photonics and with the MSI-Nucleus on Advanced Optics, Universidad de Concepción, Concepción, Chile (e-mail: gxavier@udec.cl).

G. P. Temporão and J. P. von der Weid are with the Center for Telecommunication Studies, Pontifical Catholic University of Rio de Janeiro, Rio de Janeiro, RJ, Brazil (e-mail: temporao@opto.cetuc.puc-rio.br; vdweid@opto.cetuc.puc-rio.br).
.

propagating directions between the classical and quantum signals. Note that the counter-propagation direction can be of great practical value, such as in some demonstrations of measurement device-independent QKD [16]. For the measurements, different fiber lengths of standard single-mode fiber are used up to 60 km, which is a typical mean length for a QKD fiber span [1]. These results are used in a simulation to calculate an important parameter for the performance of QKD systems: the estimated generated secure key rate as a function of the distance for a QKD system using decoy-states [17-19]. The penalty imposed by the classical communication with two different types of modulation formats is analyzed. Our results show that QKD in optical fibers populated with multiple telecom DWDM channels can be highly unfeasible unless mitigation techniques are employed [4,8,9,11].

## II. SRS Noise in the Quantum Channel

### A. Single Channel

We initially show a typical SRS spectrum, centered at the 1550 nm telecom window, generated in an optical fiber under the presence of a single classical channel. This measurement is performed with a tunable laser source (TLS) injected into the fiber at different wavelengths in the S, C, L and U-bands, while the quantum channel, fixed at 1546.12 nm was monitored [20], as depicted in Fig. 1.

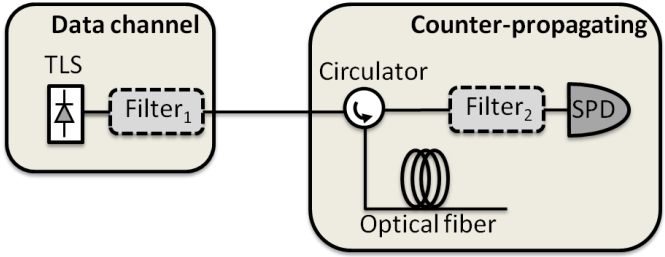

Fig. 1. Setup for evaluation of the SRS generated by a single telecom data channel, scanned over wavelength, on the fixed quantum channel. Filter$_1$ and filter$_2$ are fiber Bragg-gratings centered at the quantum channel acting as notch and bandpass, respectively.

The counter-propagating setup was deployed with a 7.5-km long standard fiber spool. The SRS noise, i.e., the counting probability at the (not populated) quantum channel, was measured for 1 mW (0 dBm) data channel power, as shown in Fig. 2. The quantum channel filtering bandwidth used was 50 GHz.

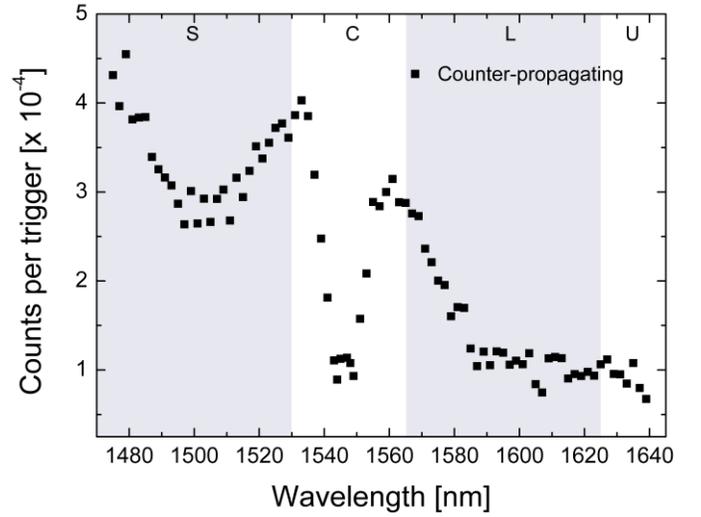

Fig. 2. Counting probability at the quantum channel (1546.12 nm with 50 GHz FWHM filtering) in a 7.5-km long standard fiber spool for counter-propagating 1 mW power of a tunable laser scanned along S, C, L and U-bands (respectively delimited by the dotted lines).

Figure 2 tells us that, in the vicinity of the quantum channel, within the telecom C-band, we can approximate the frequency dependency of the Raman counting probability as a v-shaped linear dependence. The SRS contribution to the photon-counting is given by the Raman noise power generated in the bandwidth of the filter at the detection apparatus. Hence, contribution of the data channel SRS to the number of counts per detection gate in the quantum channel can be written as [4,11,12]:

$$\begin{cases} C_{co} = P_0 z \beta(\Delta \nu) exp(-\bar{\alpha}z) \eta \frac{\tau}{h\nu} \\ C_{counter} = \frac{P_0}{2\bar{\alpha}} \beta(\Delta \nu)[1 - exp(-2\bar{\alpha}z)] \eta \frac{\tau}{h\nu} \end{cases} \quad (1)$$

where $z$ is the fiber length; $P_0$ is the launched power at the data channel; $\bar{\alpha}$ is the average fiber attenuation coefficient over the considered wavelength range of data and quantum channels, assumed to be the same; $\beta(\Delta \nu)$ is the effective SRS coefficient per unit of [km$^{-1}$] at a given frequency shift $\Delta \nu$ between data and quantum channels; $\eta$ is the efficiency of the detection apparatus, including the single-photon detector; $\tau$ is the detection gate length; and $h\nu$ is the photon energy with average frequency $\nu$.

For a given data channel frequency, the SRS contribution will be given by a Stokes or anti-Stokes coefficient, depending whether the data channel frequency is greater or smaller than the quantum channel frequency, i.e.,

$$\begin{cases} \beta_i = s(i - i_q), for\ i < i_q \\ \beta_i = a(i_q - i), for\ i > i_q \end{cases} \quad (2)$$

where $a$ and $s$ are the slopes of the frequency dependency of the SRS coefficient, $i_q$ and $i$ are the quantum channel and data channel number on the ITU-T DWDM grid.

## B. Multiple Channels

The influence of SRS generated from multiple data channels on the quantum channel is obtained by summing up the contribution of each channel with the appropriate SRS coefficient:

$$\begin{cases} C_{co}^{single} = z\,exp(-\bar{\alpha}z)\eta\frac{\tau}{h\nu}\sum_i \beta_i P_{0,i} \\ C_{counter}^{single} = \frac{1}{2\alpha}[1 - exp(-2\bar{\alpha}z)]\eta\frac{\tau}{h\nu}\sum_i \beta_i P_{0,i} \end{cases} \quad (3)$$

The counting probability due to the SRS generated was separately measured for the co- and counter-propagating cases with the setup depicted in Fig. 3.

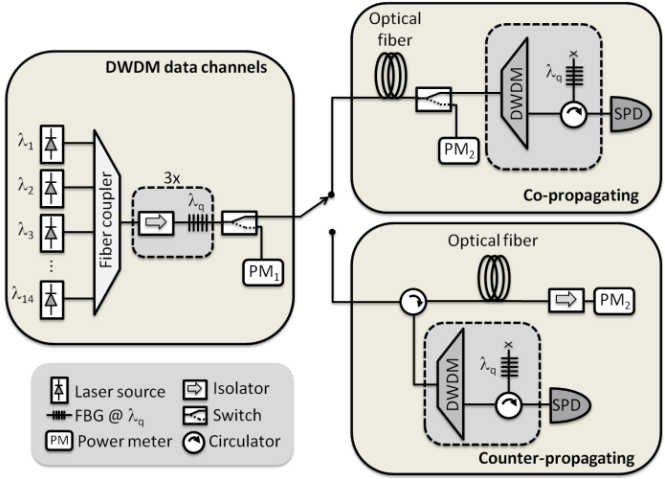

Fig. 3. Experimental setup for co- and counter-propagating signal analysis. The SRS was measured for each arrangement with different laser sources combinations and different fiber lengths. The arrow merely indicates a change in the setup for the co- and counter-propagating measurements.

A set of 14 continuous-wave (CW) distributed-feedback telecom laser diodes ($\lambda_x$), each one matching one of the ITU-T DWDM channel grid, is injected in a 16-input fiber coupler. The quantum channel under analysis, $\lambda_q$, at 1546.12 nm is devoid of any classical optical signals. The goal is to measure all the noise generated from other channels with classical power levels, as seen at the quantum channel. The laser sources pass through a set of three pairs of optical isolators and fiber-Bragg gratings (FBGs) centered at the quantum channel with 100 GHz full-width half-maximum (FWHM), in order to carve a spectral notch at this wavelength, which provides a 60 dB extinction ratio. The isolators are used to avoid the formation of optical cavities. This ratio, when combined with the 40 dB between the emission peak and the spontaneous emission of the lasers, provides a 100 dB difference between the classical and quantum signals. The employed filtering is sufficient to ensure that the contribution from crosstalk photons is negligible, which was experimentally verified by measuring the noise at the detectors with none of the fiber spools connected in the setup. The total launched optical power is measured at the power meter $PM_1$, placed after an optical switch. All laser sources are set to deliver -10.5 dBm each at $PM_1$. Table I shows the ITU-T DWDM channels used and their combination (A-G) for each measurement.

In the co-propagating setup, the DWDM channels are sent over a spool of optical fiber and then the noise scattered photons are experimentally measured in the single-photon detector (SPD). A second power meter, $PM_2$, also placed after an optical switch, is used to measure the insertion loss of the fiber spools. A filtering scheme is used to separate the quantum channel and is composed by a wavelength demultiplexer DWDM and a pair of FBGs (one with 100 GHz and the other with 10 GHz FWHM) in reflection mode (all centered at the $\lambda_q$) with a four-port circulator (simplified in Fig. 1). This provides around 60 dB extinction ratio, with an in-band insertion loss of 8.4 dB.

TABLE I
COMBINATION OF POPULATED CLASSICAL CHANNELS USED FOR THE EXPERIMENTAL MEASUREMENT OF SRS ON THE QUANTUM CHANNEL

| ITU-T CHANNEL | Frequency shift[b] [GHz] | A | B | C | D | E | F | G |
|---|---|---|---|---|---|---|---|---|
| 50 | -1110 | | | | | | | × |
| 49 | -1000 | | | | | | | × |
| 45 | -600 | | | | | × | × | × |
| 44 | -500 | | | × | × | × | × | × |
| 40 | -100 | | × | × | × | × | × | × |
| 39[a] | 0 | | | | | | | |
| 38 | 100 | × | × | × | × | × | × | × |
| 37 | 200 | × | × | × | × | × | × | × |
| 36 | 300 | | × | × | × | × | × | × |
| 35 | 400 | | | × | × | × | × | × |
| 30 | 900 | | | | | × | × | × |
| 29 | 1000 | | | | | × | × | × |
| 28 | 1100 | | | | | | × | × |
| 27 | 1200 | | | | | | × | × |
| 25 | 1400 | | | | | | × | × |

[a]Quantum channel;
[b]Relative to the quantum channel.
× Populated channels.

In the counter-propagating setup, the light is also injected in the optical fiber, but the reflected signal is routed to the filtering apparatus for analysis via an optical circulator. $PM_2$ is preceded by an optical isolator to avoid undesired back-reflections from the fiber end.

After filtering, the signal is measured, in both co- and counter-propagating cases, with an avalanche-photodiode-based SPD, operating in gated Geiger-mode. The device is set to 15% detection efficiency and opens 2.5 ns wide (effective) detection gates, internally triggered at 1 MHz with 10 $\mu s$ deadtime imposed after each detection event. The detector dark count probability is measured as $3.6 \times 10^{-5}$ per detection gate. The photon counting events are totalized during 3 intervals of 20 s for each measurement condition.

For both co- and counter-propagating setups, a set of measurements was performed with standard (SMF-28) telecom optical fibers according to the channel combinations shown in Table I. The length of the fiber spools was increased in 10 km steps up to 60 km, and the detection count rates were acquired. As previously mentioned, we also performed back-to-back measurements (0 km) to ensure that the cross-talk from the components due to an imperfect filtering was negligible and did not compromise the results with the fiber spools.

Figure 4 exhibits the ratio of counts per trigger measured for the co- and counter-propagating cases (upper and lower figures, respectively), as a function of the fiber length for the different channel combinations with 10 GHz filtering bandwidth, 2.5 ns detection gate and -10.5 dBm of optical power per channel.

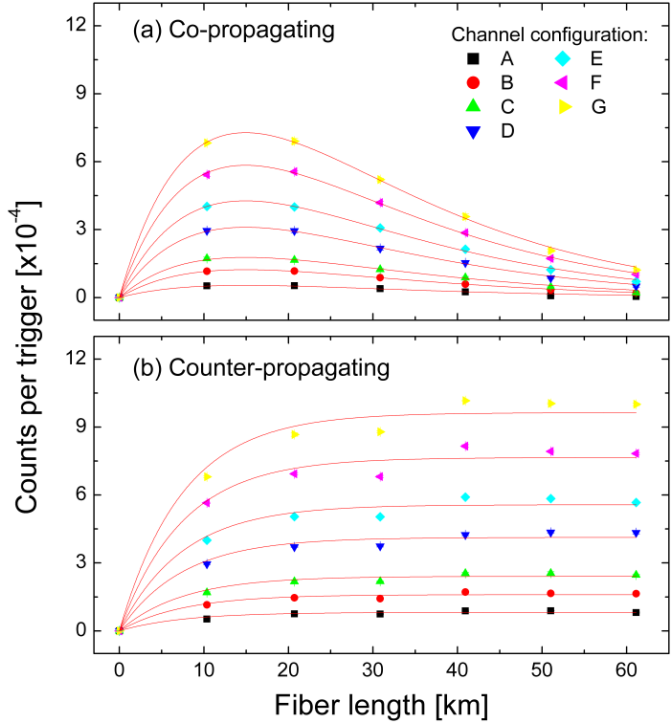

Fig. 4. Measured counting probability (symbols) with the theoretical model fit (lines) for the (a) co- and (b) counter-propagating SRS setup with channels configured as in Table I. Filtering bandwidth is 10 GHz. As expected, the channel configurations with the higher number of channels yield higher SRS detection rates.

Data in Fig. 4 is fit (solid lines) by Eq. (3). The losses in addition to the fiber attenuation were linearly considered in the SRS model. The spontaneous Raman coefficients slopes, $s$ and $a$, were extracted from experimental data of Fig. 4 and are reported (average and standard deviation) in Table II, normalized per spectral separation (in number of channels) between data and quantum channels..

TABLE II
MEASURED SPONTANEOUS RAMAN COEFFICIENT SLOPES PER CHANNEL FOR STOKES AND ANTI-STOKES SHIFTS FOR BOTH CO- AND COUNTER-PROPAGATING SETUPS

| SETUP | Co-propagating[a] [$km^{-1} \times 10^{-12}$] | Counter-propagating[a] [$km^{-1} \times 10^{-12}$] |
|---|---|---|
| Stokes | 6.9 ± 1.6[b] | 6.8 ± 1.3 |
| Anti-Stokes | 11.5 ± 1.0 | 10.8 ± 0.9 |

[a]Values per spectral separation (in number of 100 GHz channels) for 10 GHz detection bandwidth and 2.5 ns detection gate.
[b]Average ± standard deviation.

These values are used in the next section for calculating the SRS impact on a QKD system.

## C. Verification of absence of four-wave mixing noise

Figure 5 shows the measured counts per trigger as a function of the total optical power injected in the fiber for each channel configuration for co- and counter-propagating setups, as in Table I, for the arrangements described in Section IIB.

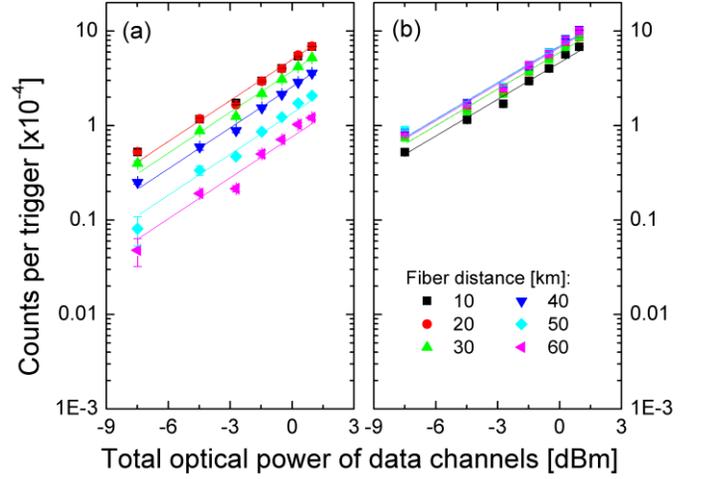

Fig. 5. Verification of the linear behavior of the scattered noise in both (a) co- and (b) counter-propagating setups as a function of the total optical power of the data channels. Lines are linear fits with unitary slopes.

The value of -10.5 dBm per data channel was verified to be low enough to avoid non-linear effects such as four-wave mixing, which could distort the SRS results. We observed a linear dependence with unitary slope of the Raman spontaneously generated detector counts as a function of increasing optical power on all 14 channels (up to a maximum of 1.5 dBm of total optical power with all laser sources enabled). This result shows that non-linear effects, such as FWM, were absent in the experimental measurements.

Additionally, the extent of the impact of FWM is also calculated for the condition of the simulations of the next section. The impact of the FWM generated by classical data channels can be neglected as long as the product $\gamma P_0 L$ is smaller than 0.1 [12], where $P_0$ is the launched power and $L$ is the fiber length. The non-linear parameter is given by

$$\gamma = n_2 \omega_0 / (c A_{eff}) \qquad (4)$$

where $A_{eff}$ is the effective area; $\omega_0$ is the angular frequency; $c$ is the velocity of light in vacuum; and $n_2$ is the nonlinear parameter of the fiber related to $\chi^{(3)}$.

Considering $A_{eff}$ of 50 μm² and $n_2$ of 2.6×10⁻²⁰ m/W [21], the FWM can be estimated for SMF-28 telecom optical fibers near the zero dispersion wavelength around 1310 nm. In our experiment, we operate far from this wavelength, in the ITU-T C-band. A phase-match efficiency factor ($\eta_{FWM}$) can then be introduced as the FWM generation is not favored in such a condition.

Following [22], we first estimate the propagation constant difference for three DWDM wavelength channels (37, 38 and 39) as

$$\Delta k = \frac{2\pi\lambda_k}{c}\left(\Delta f_{eq}\right)^2\left[D_c + \frac{\lambda_k^2}{2}\Delta f_{eq}\frac{dD_c}{d\lambda}\right] \quad (5)$$

being $\Delta f_{eq} = \left(\Delta f_{ik}\Delta f_{jk}\right)^{1/2}$ and $\Delta f_{mn} = |f_m - f_n|$, with $m, n = i, j, k$. The dispersion parameter $D_c$ is 16 ps km$^{-1}$nm$^{-1}$ at 1550 nm and zero at 1310 nm. Its slope is 66.7 s/m for anomalous dispersion.

The FWM efficiency can be calculated, based on the phase mismatch as

$$\eta_{FWM} = \frac{\alpha^2}{\alpha^2+\Delta k^2}\left[1 + \frac{4e^{-\alpha L}\sin^2(\Delta kL)}{\left(1-e^{-\alpha L}\right)^2}\right] \quad (6)$$

The ratio between the efficiency obtained for phase matched and non-phase matched condition described is of 2.2×10$^{-5}$. The poor FWM generation efficiency when all channels are far from the zero dispersion wavelength allows for the operation of many data channels with usual power levels without jeopardizing the quantum communication with FWM effects.

### III. IMPACT OF SRS ON QKD SYSTEMS

The analysis of the impact of SRS generated by classical traffic on QKD systems is based on the model presented in [18,19] for a BB84 system with a faint laser source with decoy states. This means that the intensity of the quantum source is assumed to be randomly varied to test the channel against the intervention of an eavesdropper. It enables reaching longer distances and the analysis results in a secure key rate value, lower bounded by the model. We conservatively assume here the usage of infinite decoy states and the analysis depends on the average number of emitted signal photons per time interval, on the channel loss, on the in-band generated SRS, and on the detection apparatus efficiency.

The channel transmittance between Alice and Bob depends on the fiber length $L$ and is given by

$$\eta = exp(-\alpha L) \times \eta_{bob} \times \eta_{spd} \quad (7)$$

where $\alpha$ is the fiber attenuation coefficient at the quantum channel [$km^{-1}$], $\eta_{bob}$ is the transmittance of Bob's equipment and $\eta_{spd}$ is the SPD detection efficiency. The yield of a number state containing *n* photons, defined as the conditional probability of a detection event at Bob's side given that Alice sends out an *n*-photon state [18], can be written, for small $\eta Y_0$ values, as

$$Y_n \approx Y_0 + 1 - (1-\eta)^n \quad (8)$$

where the yield for vacuum state ($Y_0$) is the system noise, composed by the SRS noise and the dark counts of the SPDs. For a BB84 system [1] with two SPDs, $Y_0 = 2P_{dark} + \kappa P_{srs}$, i.e., twice the dark counts probability ($P_{dark}$) of one detector, and the unpolarized SRS noise $P_{srs}$ split between the two devices. The factor $\kappa$ depends on the modulation format assumed for the telecom traffic and is related to its duty cycle.

For *phase shift-keying*-based (PSK) format, this factor is 1, while for *on-off keying* (OOK) with *return to zero* (RZ), $\kappa$ is ¼. The PSK-based format family represents the worst case, when intensity of the parallel traffic is continuous in time, while in the RZ-OOK, half the bits are zero and the bits occupy half of each bit slot, which results in a smaller average optical power and a linear reduction of the SRS.

A faint laser source can be approximated to a weak coherent state, which exhibits Poisson distribution of the number of photons *n* per time interval, i.e., P(n|μ) = exp(−μ)μ$^n$/n!, given an average number of *μ* photons over each interval. The channel gain $Q$ is composed by the sum over the gain values for each photon-number state, i.e., the probability that Bob yields a detection event given that Alice has sent an n-photons pulse,

$$Q = \sum_{n=0}^{\infty} Y_n P(n|\mu) = Y_0 + 1 - exp(-\mu\eta) \quad (9)$$

Since each vacuum yield can randomly generate a count in any detector, the associated quantum bit error rate (QBER) is ½. Furthermore, misalignment of the optical components ($\gamma$) can bring non-vacuum states to cause erroneous detection. The overall QBER $E$ can be calculated by summing over the contribution of each photon-number state relative to the overall gain, i.e.,

$$E = \frac{½Y_0 + \gamma[1-exp(-\mu\eta)]}{Q} \quad (10)$$

The lower bound for the final secure key rate is finally written as [19]

$$R > \frac{1}{2}\left\{Q_1\left[1 - H_2\left(\frac{½Y_0+\gamma\eta}{Y_1}\right)\right] - Qf(E)H_2(E)\right\} \quad (11)$$

and depends on the single-photon gain $Q_1$ and the Shannon entropy of the error $H_2(E)$. A quantum error correction inefficiency factor $f(E)$ is usually considered, which reduces the final key rate.

#### A. Results

The data presented in [23] was used for the simulations, with the exception of the spontaneous Raman noise, which was continuously calculated as a function of transmission distance for the channel configurations of Table I with the coefficients from Table II. The detection efficiency is $\eta_{bob} \times \eta_{spd} = 0.045$, with dark count probability of $0.85 \times 10^{-6}$ per SPD, and the transmittance of the fiber link is calculated for each fiber length with $\bar{\alpha} = 0.0484\ km^{-1}$. The average signal photon number per time interval is 0.50 and the optical misalignment factor $\gamma$ is 0.033 [23]. The quantum error correction inefficiency factor is obtained from [24].

The vacuum yield is calculated with $P_d$ and our experimental value of $P_{srs}$ obtained for each channel combination from Table I with the spontaneous Raman coefficient of Table II using 0 dBm to each channel. This power adjustment can be applied directly since SRS scales linearly with optical power. The filtering bandwidth was varied and the detection gate window is 1 ns.

Figure 6 shows the secret key rate generated according to the link distance between Alice and Bob when classical telecom data is sent through the same fiber co- and counter propagating, (a,b) and (c,d) respectively. The filtering bandwidth was set to 100 and 10 GHz and data was assumed to be PSK modulated ($\kappa = 1$).

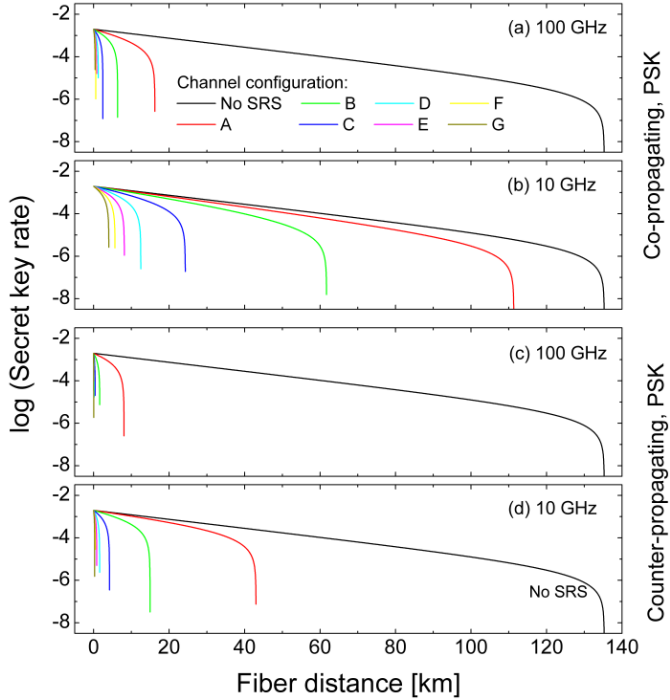

Fig. 6. Secret key rate as a function of the fiber link length when PSK-modulated telecom traffic is sent with 1 mW per channel, (a,b) co- and (c,d) counter-propagating relative to the quantum channel. The filtering bandwidth is (a,c) 100 and (b,d) 10 GHz FWHM.

PSK-based modulation represents the worst-case when classical and quantum signals share the same optical fiber. This always-on kind of modulation results in full energy transfer from data channels to the quantum channel, as all telecom channels keep constant power levels over time. A better scenario would be to use OOK modulation, specially the RZ format. This means that each bit does not occupy the whole bit slot and the average power sent into the quantum channel is averaged over the duty cycle. Of course our assumption does not consider any kind of synchronization between quantum and classical signals (as a time-domain multiplexing scheme) [8], which could reduce the SRS impact on a QKD system, which generates an extra cost due to increased network complexity. Figure 7 shows the secret key rate as a function of the fiber link distance when the co-existent classical telecom data is OOK-RZ-modulated and sent co- and counter propagating, (a-c) and (d-f) respectively. The filtering bandwidth was again set to 100 and 10 GHz.

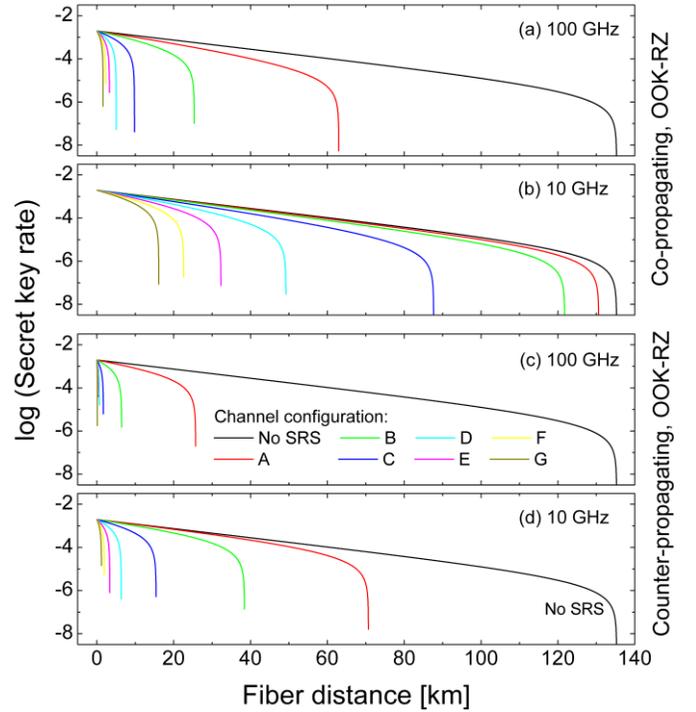

Fig. 7. Secret key rate as a function of the fiber link length when OOK-RZ-modulated telecom traffic is sent with 1 mW per channel, (a,b) co- and (c,d) counter-propagating relative to the quantum channel. The filtering bandwidth is (a,c) 100 and (b,d) 10 GHz FWHM.

The figures also depict the ideal case, with no SRS generation. Telecom traffic sent alongside the quantum channel may impose a harsh penalty on the secret key rate and, consequently, in the link distance. The impact of the scattered noise depends on the number of classical channels and their relative spectral position to the quantum channel and also on their power value. The effective noise measured on the quantum channel is the integrated spectral power density over the filtering bandwidth and its impact is reduced by narrower filtering.

The maximum achievable distance that results in a positive secret key rate was computed for each channel configuration reported in Table II. Since the contribution for the SRS changes linearly with data channels power [4,11,20], we performed another set of simulations to determine the maximum achievable distance as a function of this power. The power for each classical telecom channel was varied from -10 to 0 dBm and different filtering bandwidths (100 and 10 GHz FWHM) were considered. The range for the optical power was chosen in order to provide the reader a broad overview of the scattered noise. The range is upper bounded by the usually employed optical power for data channels, while the lower bound could impact on the bandwidth-distance compromise of the classical link. In some configurations, higher optical power values can generate other non-linear effects, like FWM, specially when operation is near the zero dispersion wavelength of the fiber. The results for the co- and counter propagating cases with PSK-based modulation are shown in Fig. 8.

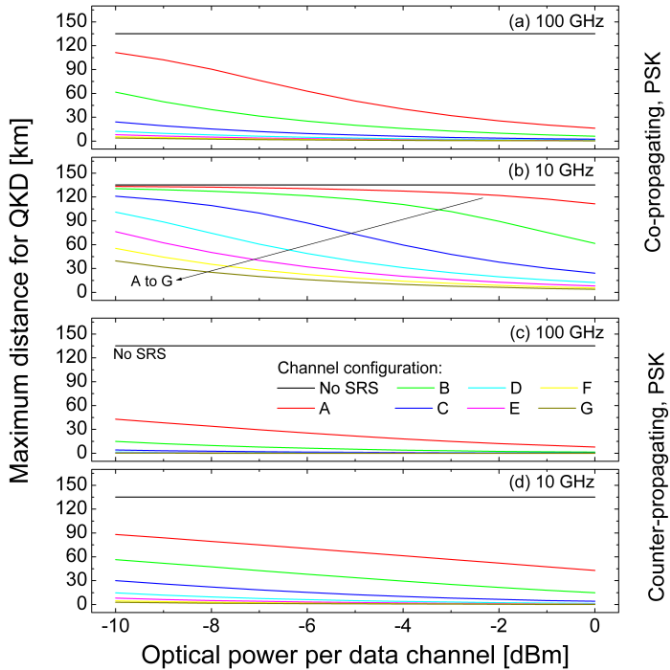

Fig. 8. Maximum QKD achievable distance as a function of the classical telecom channels power (per channel). Classical data is PSK-modulated and (a,b) co- or (c,d) counter-propagates relative to the quantum channel. The filtering bandwidth is (a,c) 100 and (b,d) 10 GHz FWHM.

Similar results were generated for the OOK-RZ modulation format. The co- and counter-propagating cases are depicted in Fig. 9.

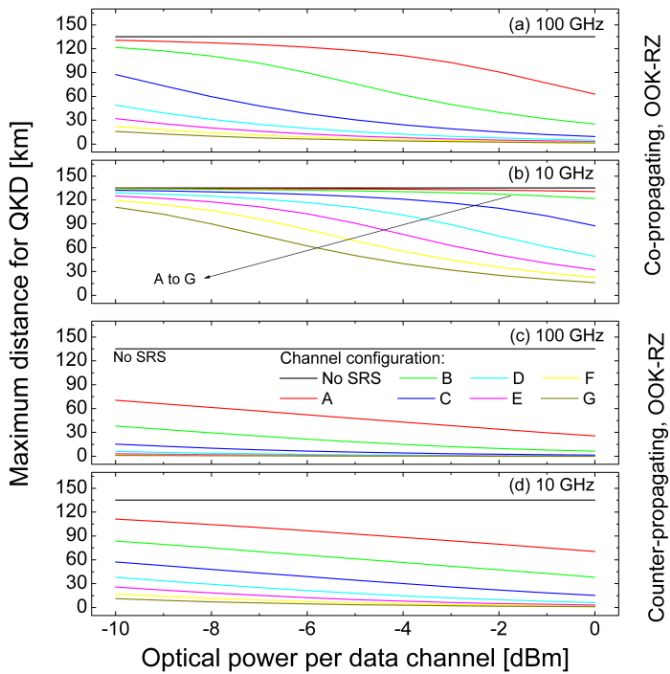

Fig. 9. Maximum QKD achievable distance as a function of the classical telecom channels power (per channel). Classical data is OOK-RZ-modulated and (a-c) co- or (d-f) counter-propagates relative to the quantum channel. The filtering bandwidth is (a,d) 100, (b,e) 10 and (c,f) 1 GHz FWHM.

A smaller filtering bandwidth can linearly reduce the generation of Raman scattered noise. However, care must be taken regarding the bandwidth of the single-photon pulses. Depending on the frequency employed, some attenuation of the quantum signal can be imposed by the filter, due to the Fourier-transform limit [14].

For high data channel power levels other non-linear effects, as four-wave mixing, may arise when using multiple channels.

## IV. CONCLUSION

We present experimental results showing the impact of the SRS noise generated from multiple telecom optical signals on a single QKD channel, all co-existing in the ITU-T C-band in the same fiber. The impact of the scattered noise on the quantum channel was experimentally evaluated. The final secret key rate was calculated under the presence of spontaneous Raman scattering for different fiber link lengths in both co- and counter-propagating setup with different number of DWDM channels present. Different modulation formats for the classical communication were compared, with a rate/distance penalty depending on the traffic duty cycle and on the quantum channel filtering bandwidth. The maximum achievable distance with positive secret key rate was estimated, with more favorable results when quantum and classical communication co-propagate. Furthermore it has been shown that the best configuration is to populate the telecom channels as close as possible from the quantum one, preferably positioned at longer wavelengths to avoid the higher Stokes shift peak contribution of the shorter wavelengths. Additionally, the operation far from the zero dispersion of the fiber greatly reduces the phase-matching condition for generation of four-wave mixing inside the quantum channel. An additional possibility is to reduce the detection bandwidth as much as possible, linearly reducing the Raman noise and extending the maximum achievable QKD distance.